\documentclass{elsart}
\usepackage{amsmath}
\usepackage[hiresbb,pdftex]{graphicx}
\usepackage{amssymb}
\usepackage{acronym}

\input{tcilatex}

\begin{document}

\begin{frontmatter}
\title{The Spread of Infectious Disease with Household-Structure on the Complex Networks}

\author{\corauthref{cor1}Jingzhou Liu$^{1}$}
\author{,Jinshan Wu$^{2}$}
\author{ and Z.R. Yang$^{1}$}

\address{$^{1}$Department of physics, Beijing Normal University,
Beijing 100875, China } 
\address{$^{2}$Department of Physics, Simon Fraser University, Burnaby, B.C., Canada, V5A 1S6 } 
\corauth[cor1]{Corresponding author. E-mail address: liujz606@hotmail.com}
\begin{abstract}\ \  In this paper we study the household-structure SIS epidemic spreading on general
complex networks. The household structure gives us the way to distinguish
inner and the outer infection rate. Unlike household-structure models on homogenous
networks, such as regular and random networks, here we consider heterogeneous
networks with arbitrary degree distribution p(k). First we introduce the epidemic model. Then rate equations under
mean field appropriation and computer simulations are used here to analyze our
model. Some unique phenomena only existing in divergent network with household
structure is found, while we also get some similar conclusions that some simple geometrical
quantities of networks have important impression on infection property
of infectous disease. It seems that in our model even when local cure rate is greater than inner infection rate in every household,
disease still can spread on scale-free network. It implies that no disease is spreading
in every single household, but for the whole network, disease is spreading. Since
our society network seems like this structure, maybe this conclusion remind us that
during disease spreading we should pay more attention on network structure than
local cure condition.
\end{abstract}
\begin{keyword}Infectious Disease; SIS Model; Networks
\PACS 89.75.-Hc; 05.70.Ln; 02.10.Yn; 87.23.Ge; 64.70.-p
\end{keyword}
\end{frontmatter}

\section{Introduction}

The spread of disease has been one of the focuses in the field of
statistical physics for many years. The dynamical behavior of so-called
susceptible-infected-susceptible (SIS) model and susceptible- infected-
removed (SIR) model have been widely investigated on regular network and
complex networks[1-12]. Within the studying, individuals are modeled as
sites and possible contacts between individuals are linked by edges between
the sites. It is easy to see that both the properties of disease and
topological character of network determine the dynamics of the spread of
disease. Studies have showed that there is an epidemic threshold $\lambda
_{c}$ on regular networks. If the effective spreading rate $\lambda $%
\TEXTsymbol{>}$\lambda _{c}$, the infection spreads and becomes endemic;
otherwise the infection will die out. While the threshold disappears on
scale-free networks[4].

Usually, infectious diseases, such as HIV and computer virus, have the
similar spreading property. They not only can spread in one household, but
also can spread from one household to another. To study this spreading
character, there have been of considerable interests to epidemic models
spreading among a community of households[12-17]. These studies were
concerned with SIR model, which cannot appear endemic behavior.\ In 1999,
Ball introduced the SIS household-structure model[18], in which the
population is partitioned into $m$ households with $N$ members in each
household. A threshold parameter $R_{\ast }$ was defined. It is shown that
for the household with $2$ members, if $R_{\ast }<1$ then the epidemic die
out; if $R_{\ast }>1$ the epidemic will exist at an endemic equilibrium.\
This model has also been studied on homogeneous network by the mean of
self-consistent field[19,20]. The similar results have been obtained. These
previous studies about household-structure epidemic model were mainly on
regular networks. However, studies have showed that a large number of
systems, such as Internet, world-wide-web, physical, biological, and social
networks, exhibit complex topological properties[21-23]. In particular,
small-world properties[24] and scale-free degree distributions[25] appear in
many real network systems. In this paper, we will analyze the SIS
household-structure epidemic model on complex networks. The outline is as
follows: 1) introduction; 2) description of the model; 3) mean-field
equations; 4) steady-state solutions; 5) simulation; 6) summary.

\section{Model}

In complex networks with degree distribution $p(k)$, which is the
probability that a given site has $k$ connections (links) that connect it
with other $k$ sites (We say that the given site' degree is $k$.), there are 
$N$ individuals that are grouped as a household on every site. We assume
that these N individuals contact each other fully. A healthy individual may
get infected from within the household and from outside its household. The
parameters $\lambda $ and $\beta $ are the infection rates from outside and
from within the household respectively. We give each site $x$ a number $%
i(i\in \lbrack 0,N])$, which means that there are $i$ infected individuals
in the household at site $x$. The number of infected individuals at a given
site $x$ changes according to the following transition rates:\bigskip

$\ \ \ \ \ \ \ \ \ \ \ \ \ \ \ \ \ \ \ \ \ \ 0\longrightarrow 1$ at rate $%
\lambda \sum_{\left\langle x,y\right\rangle }i_{y}\medskip $

\ \ \ \ \ \ \ \ \ \ \ \ \ \ \ \ \ \ \ \ \ \ $i\longrightarrow i+1$ at rate $%
i\beta $ for $1\leq i\leq N-1\medskip $

\ \ \ \ \ \ \ \ \ \ \ \ \ \ \ \ \ \ \ \ \ \ $i+1\longrightarrow i$ at rate $%
\gamma $ for $0\leq i\leq N-1\medskip $

In the above expressions $\left\langle x,y\right\rangle $means that site $x$
and site $y$ are nearest neighbors, and we suppose there is a (connection)
link between them. Infected individuals may infect healthy individuals in
their household with rate $\beta ,$and also can infect healthy individuals
in their nearest neighbors with rate $\lambda .$We assume that once a site
is infected, infections within the site are much more likely than infections
from outside, so we can neglect the latter. And also, an infected individual
in a site can recover with rate $\gamma .$We suppose that all the
individuals in a household have the same external connectivity and do not
take the birth and death into account.

\section{Mean-field equations}

We now solve the above model with mean-field method. Let $u_{k,i}$ be the
density of individuals whose household has $i$ infected individuals and the
corresponding site's degree is $k$, which means this site has $k$ nearest
neighbors. It is worth noticing that $\sum_{i=0}^{N}u_{k,i}=1$. According to
the transitions rate described in the above\ section, the evolution
equations of $u_{k,i}$are written as below[4,6]:

\begin{eqnarray}
\frac{\partial u_{k,0}(t)}{\partial t} &=&\gamma u_{k,1}-\lambda k\Theta
_{k}(t)u_{k,0}\text{ \ \ \ \ \ \ \ \ \ \ \ \ \ \ \ \ \ \ \ \ \ \ \ \ \ \ \ }
\label{g1} \\
\frac{\partial u_{k,1}(t)}{\partial t} &=&\lambda k\Theta
_{k}(t)u_{k,0}-\beta u_{k,1}+\gamma u_{k,2}-\gamma u_{k,1}\text{ \ \ \ \ \ \
\ }  \label{g2} \\
\frac{\partial u_{k,i}(t)}{\partial t} &=&(i-1)\beta u_{k,i-1}-i\beta u_{k,i}%
\text{\ +}\gamma u_{k,i+1}-\gamma u_{k,i}\text{ \ }(i\in \lbrack 2,N-1])
\label{g3} \\
\frac{\partial u_{k,N}(t)}{\partial t} &=&(N-1)\beta u_{k,N-1}-\gamma u_{k,N}
\label{g4}
\end{eqnarray}

In equations(\ref{g1})-(\ref{g4}), $\Theta _{k}(t)$ is the probability that
a link from a site points to another site with at least one infected
individual. And the expression of $\Theta _{k}(t)$ is:

\begin{equation}
\Theta _{k}(t)=\sum_{k^{^{\prime }}}p(k^{^{\prime
}}/k)\sum_{j=1}^{N}ju_{k^{^{\prime }},j}(t)  \label{theta}
\end{equation}%
where $p(k^{^{\prime }}/k)$ is the probability that a site with $k$ degrees
points to another site with $k^{^{\prime }}$ degrees. For uncorrelated
networks the expression of $p(k^{^{\prime }}/k)$ is[7]:%
\begin{equation}
p(k^{^{\prime }}/k)=\frac{k^{^{\prime }}p(k^{^{\prime }})}{\sum_{k^{^{\prime
}}}k^{^{\prime }}p(k^{^{\prime }})}=\frac{1}{\left\langle k\right\rangle }%
k^{^{\prime }}p(k^{^{\prime }})  \label{pk}
\end{equation}

Substituting (\ref{pk}) to (\ref{theta}), we get $\Theta _{k}(t)=\Theta (t)$
independent of $k$:%
\begin{equation}
\Theta _{k}(t)=\Theta (t)=\frac{1}{\left\langle k\right\rangle }%
\sum_{k^{^{\prime }}}k^{^{\prime }}p(k^{^{\prime
}})\sum_{j=1}^{N}ju_{k^{^{\prime }},j}(t)  \label{theta2}
\end{equation}

Now we are going to get steady solutions of Eqs.(\ref{g1})-(\ref{g4})

\section{Steady-state solutions}

\subsection{For N=2}

The evolution equations are simplified as:

\begin{eqnarray}
\frac{\partial u_{k,1}(t)}{\partial t} &=&\lambda k\Theta
(t)(1-u_{k,1}-u_{k,2})+\gamma u_{k,2}-\beta u_{k,1}-\gamma u_{k,1}
\label{g9} \\
\frac{\partial u_{k,2}(t)}{\partial t} &=&\beta u_{k,1}-\gamma u_{k,2}
\label{g7}
\end{eqnarray}

In Eq.(\ref{g9}), we have used the equality: $u_{k,0}+u_{k,1}+u_{k,2}=1$.
When $t\rightarrow \infty $, $\frac{\partial u_{k,1}}{\partial t}=0$ and $%
\frac{\partial u_{k,2}}{\partial t}=0$, then the steady-state solutions are:%
\begin{eqnarray}
u_{k,1} &=&\frac{\lambda \gamma k\Theta }{\gamma ^{2}+(\gamma +\beta
)\lambda k\Theta }  \label{u21} \\
u_{k,2} &=&\frac{\lambda \beta k\Theta }{\gamma ^{2}+(\gamma +\beta )\lambda
k\Theta }  \label{u22}
\end{eqnarray}

Substituting (\ref{u21})and (\ref{u22}) to (\ref{theta2}), we get the
self-consistent equation of $\Theta $:%
\begin{equation}
\Theta =\frac{1}{\left\langle k\right\rangle }%
\sum_{k}kp(k)(u_{k,1}+2u_{k,2})=\frac{1}{\left\langle k\right\rangle }%
\sum_{k}kp(k)\frac{(\gamma +2\beta )\lambda k\Theta }{\gamma ^{2}+(\gamma
+\beta )\lambda k\Theta }  \label{theta3}
\end{equation}

Clearly, $\Theta =0$ is a solution of Eq.(\ref{theta3}), which implies that $%
u_{k,0}=1$, $u_{k,1}=0$ and $u_{k,2}=0$ is a steady-state solution of Eqs.(%
\ref{g9}) and (\ref{g7}). A nonzero steady-state solution $\Theta $ (That
is: $u_{k,i}\neq 0$, for $i>0$) is obtained when $\gamma $, $\beta $ and $%
\lambda $ satisfy the following inequality:

\begin{equation*}
\frac{1}{\left\langle k\right\rangle }\sum_{k}kp(k)\frac{(\gamma +2\beta
)\lambda k}{\gamma ^{2}}\geq 1
\end{equation*}

Then we can get the spreading threshold:%
\begin{equation}
\lambda _{c}=\frac{\gamma ^{2}}{\gamma +2\beta }\frac{\left\langle
k\right\rangle }{\left\langle k^{2}\right\rangle }  \label{N2c}
\end{equation}%
where $\left\langle k\right\rangle =\sum_{k}kp(k)$, $\left\langle
k^{2}\right\rangle =\sum_{k}k^{2}p(k)$. In other words, the disease will die
out when $\lambda <\lambda _{c}$; otherwise the disease will pervade the
system. Clearly, the threshold $\lambda _{c}$ is the function of $\beta $, $%
\gamma $ and $\frac{<k>}{<k^{2}>}$. So the degree distribution of networks
plays an important role on $\lambda _{c}$.

For $p(k)=\delta _{k,k_{c}}$, the network is homogeneous and $\lambda _{c}=%
\frac{\gamma ^{2}}{\gamma +2\beta }\frac{1}{k_{c}}$. We can lift $\lambda _{c%
\text{ }}$to prevent infection in terms of increasing the recover rate $%
\gamma $ or decreasing the site degree $k_{c}$.

For $p(k)=Ck^{-\upsilon }$ $(\upsilon \in (2,3])$, the networks are
scale-free[19]. When $k\rightarrow \infty $, $\frac{\left\langle
k\right\rangle }{\left\langle k^{2}\right\rangle }\rightarrow 0$, the
threshold is absent. This fact implies that for any positive value of $%
\lambda $ the infection can pervade the system, which is the same as the
standard SIS model[4].

\subsection{N\TEXTsymbol{>}2}

Let $\frac{\partial u_{k,i}}{\partial t}=0(i=1,2,\cdots N)$. Suppose $%
\mathbf{U}_{k}\mathbf{=}(u_{k,1},u_{k,2},\cdots ,u_{k,N})^{\mathbf{T}}$ and $%
\mathbf{V=}(1,0,\cdots ,0)^{\mathbf{T}}$. Considering $%
\sum_{j=0}^{N}u_{k,j}=1,$then Eqs.(\ref{g1})-(\ref{g4}) can be written as:%
\begin{equation}
\mathbf{SU}_{k}\mathbf{=}-\lambda k\Theta \mathbf{V}  \label{g13}
\end{equation}

The matrix \textbf{S }is:

\ \ \ \ \ \ \ \ \ 
\begin{equation}
\mathbf{S=}\left( 
\begin{array}{cccccc}
-\gamma -\beta -\lambda k\Theta & \gamma -\lambda k\Theta & -\lambda k\Theta
& \cdots & -\lambda k\Theta & -\lambda k\Theta \\ 
\beta & -2\beta -\gamma & \gamma & \cdots & 0 & 0 \\ 
0 & 2\beta & -3\beta -\gamma & \cdots & 0 & 0 \\ 
\vdots & \vdots & \vdots & \cdots & \vdots & \vdots \\ 
0 & 0 & 0 & \cdots & -(N-1)\beta -\gamma & \gamma \\ 
0 & 0 & 0 & \cdots & (N-1)\beta & -\gamma%
\end{array}%
\right)  \label{matrixs}
\end{equation}

Since $\det (\mathbf{S})=(-\gamma )^{N}-\lambda k\Theta
\dsum\nolimits_{j=1}^{N}(j-1)!(-\gamma )^{N-j}(-\beta )^{j-1}\neq 0$, so $%
\mathbf{S}^{-1}$ exists. Thus:%
\begin{equation}
\mathbf{U}_{k}=-\lambda k\Theta \mathbf{S}^{-1}\mathbf{V}  \label{g15}
\end{equation}

\begin{equation}
\sum_{j=1}^{N}ju_{k,j}=\mathbf{n\cdot U}_{k}=(-\lambda k\Theta )\mathbf{nS}%
^{-1}\mathbf{V=-\lambda }k\Theta \mathbf{n(S}^{-1}\mathbf{V)=-\lambda }%
k\Theta \sum_{j=1}^{N}\mathbf{n}_{j}\mathbf{S}_{j1}^{-1}  \label{g16}
\end{equation}

where $\mathbf{n=}(1,2,\cdots ,N)$ and $\mathbf{n}_{j}=j$

From (\ref{matrixs}), we get $\mathbf{S}_{j1}^{-1}$:%
\begin{equation}
\mathbf{S}_{j1}^{-1}=\frac{(-\gamma )^{N-j}(-\beta )^{j-1}(j-1)!}{(-\gamma
)^{N}-\lambda k\Theta \dsum\nolimits_{j=1}^{N}(j-1)!(-\gamma )^{N-j}(-\beta
)^{j-1}}  \label{g17}
\end{equation}

Substituting (\ref{g17}) and (\ref{g16}) to (\ref{theta2}), we get the
self-consistent equation of $\Theta $:%
\begin{equation}
\Theta =-\frac{1}{\left\langle k\right\rangle }\sum_{k}\sum_{j}\frac{\lambda
k^{2}p(k)(-\gamma )^{N-j}(-\beta )^{j-1}j!\Theta }{(-\gamma )^{N}-\lambda
k\Theta \dsum\nolimits_{j=1}^{N}(j-1)!(-\gamma )^{N-j}(-\beta )^{j-1}}
\end{equation}

That is:%
\begin{equation}
\Theta =-\frac{1}{\left\langle k\right\rangle }\left\langle \sum_{j}\frac{%
\lambda k^{2}(-\gamma )^{N-j}(-\beta )^{j-1}j!\Theta }{(-\gamma
)^{N}-\lambda k\Theta \dsum\nolimits_{j=1}^{N}(j-1)!(-\gamma )^{N-j}(-\beta
)^{j-1}}\right\rangle   \label{g18}
\end{equation}

Obviously, $\Theta =0$ is a solution of Eq.(\ref{g18}). In addition, a
non-zero solution with $\Theta \neq 0$ and $u_{k,i}\neq 0$ $(i=1,2,\cdots ,N)
$ is allowed if the following inequality holds:%
\begin{equation}
\left( -\frac{1}{\left\langle k\right\rangle }\left\langle \sum_{j}\frac{%
\lambda k^{2}(-\gamma )^{N-j}(-\beta )^{j-1}j!}{(-\gamma )^{N}}\right\rangle
\right) \geq 1  \label{g19}
\end{equation}

That is:%
\begin{equation}
\lambda \frac{\left\langle k^{2}\right\rangle }{\left\langle k\right\rangle }%
\sum_{j=1}^{N}\frac{1}{\gamma }\left( \frac{\beta }{\gamma }\right)
^{j-1}j!\geq 1  \label{g20}
\end{equation}

From (\ref{g20}), we get the epidemic threshold:%
\begin{equation}
\lambda _{c}=\frac{1}{f(N,\beta ,\gamma )}\frac{\left\langle k\right\rangle 
}{\left\langle k^{2}\right\rangle }  \label{g21}
\end{equation}%
where $f(N,\beta ,\gamma )=\sum_{j=1}^{N}\frac{1}{\gamma }\left( \frac{\beta 
}{\gamma }\right) ^{j-1}j!$, and $f(N,\beta ,\gamma )$ is an increasing
function of $N$ and $\gamma $, but a decreasing function of the recover rate 
$\gamma $. So the epidemic threshold is determined by three parameters$%
(N,\beta ,\gamma )$ and the networks degree distribution $p(k)$. We notice
that the expression (\ref{g21}) involves multiplication of the well-known
term $\frac{\left\langle k\right\rangle }{\left\langle k^{2}\right\rangle }$%
[2,4,6,9], which is closely related to the "average" number of secondary
infections[7,8]. Not surprising, this result is the same as that of the
standard SIS model[4].

For $p(k)=\delta _{k,k_{c}}$, the network is homogeneous. Then $\lambda _{c}=%
\frac{1}{f(N,\beta ,\gamma )}\frac{1}{k_{c}}$, we can increase the recover
rate $\gamma $ or decrease the site degree $k_{c}$ and the size of the
household $N$ to lift $\lambda _{c\text{ }}$to prevent the infectious
disease from spreading. For large $N$ the threshold is very small.

For $p(k)=Ck^{-\upsilon }$ $(\upsilon \in (2,3])$, the network is
scale-free[21]. When $k\rightarrow \infty $, $\frac{\left\langle
k\right\rangle }{\left\langle k^{2}\right\rangle }\rightarrow 0$, then $%
\lambda _{c}=\frac{1}{f(N,\beta ,\gamma )}\frac{\left\langle k\right\rangle 
}{\left\langle k^{2}\right\rangle }\rightarrow 0$. So the threshold is
absent for scale-free network. This implies that for any positive value of $%
\lambda $, the infection can pervade the system even with high recover rate.

\section{Simulation result}

In above section, we have given the analytical result of the SIS model with
household structure. We find that for regular network there is an epidemic
threshold $\lambda _{c}$; while for scale-free network the threshold
disappears. For comparison, we simulate the model on regular network(see Fig%
\ref{fig1}) and on scale free network(see Fig\ref{fig2}) respectively. For
simplicity(without lack of generality), we set $\gamma =1$, $N=4$. In Fig.1,
we plot the fraction of infected individuals in the stationary state, $\rho $%
, for different values of $\beta $ on regular network with $k_{c}=4$.
Obviously, there is a threshold $\lambda _{c}$ for each $\beta $. For $\beta
=0.6$, $\lambda _{c}$ is $0.026$, in agreement with the corresponding
analytical result, $\lambda _{c}=0.026$, which can be obtained from(\ref{g21}%
). Only when $\lambda $ is increased above $\lambda _{c}$ is a significant
prevalence observed. In Fig.2, we plot the fraction of infected individuals
in the stationary state, $\rho $, for different values of $\beta $ on
scale-free network with $\left\langle k\right\rangle =6$. We observe that$\
\lambda _{c}$ is absent. In contract with the standard SIS model, of which
the prevalence, $\rho $, increases slowly when increasing $\lambda $[24],
our current epidemic model exhibits that $\rho $ increases rapidly with $%
\lambda $.

\section{Summary}

In this work, we analyze the SIS model that incorporates social household.
We have focused on the impaction of geometrical property of complex networks
and on the role of several parameters in the spreading threshold. Results
show that the large household size N and the high within household infection
rate are more likely to cause the spread of disease. But it's worth noticing
that, even when local recovery rate is greater than effective infection
rate, in divergent networks such as scale-free network, disease still can
spread! This results tell us that even the local recover condition is good
enough to give local protection, there are still some probability for a wide
range disease spreading. It seems that this phenomenon can only exist in
divergent networks with household structure. Maybe this imply that we have
to care about the network structure much more than recover condition during
disease spreading.

Of course, the model we have studied seems more ideal. For example, we have
supposed that the existence of the N-member households do not affect the
property of the complex networks, and also we do not take the move of the
individuals into account. However, the result tells us that the properties
of the complex networks play the most important role in the epidemic
spreading.

\begin{center}
ACKNOWLEDGMENT
\end{center}

This work was supported by the National Science Foundation of China under
Grant No. 10175008. We thank research professor Yifa Tang for helpful
discussion. We also acknowledge the support from The State Key Laborary of
Scientific and Engineering Computering (LSEC), Chinese Academic of Science.

\begin{center}
REFERENCE
\end{center}

[1]R. M. May and A. L. Lloyd, Phys. Rev. E \textbf{64}, 066112 (2001).

[2]R. Pastor-Satorras and A. Vespignani, Phys. Rev. E \textbf{65}, 035108(R)
(2002).

[3]M. E. J. Newman, Phys. Rev. E \textbf{66}, 016128 (2002).

[4]R. Pastor-Satorras and A. Vespignani, Phys. Rev. Lett. \textbf{86}, 3200
(2001); Phys. Rev. E \textbf{63}, 066117 (2001).

[5]Y. Moreno, J. B. G\'{o}mez, and A. F. Pacheco, Phys. Rev. E \textbf{68},
035103 (2003).

[6]M. Bogu\~{n}\'{a} and R. Pastor-Satorras, Phys. Rev. E \textbf{66},
047104 (2002).

[7]M. Bogu\~{n}\'{a}, R. Pastor-Satorras, and A. Vespignani, Phys. Rev.
Lett. \textbf{90}, 028701 (2003).

[8]D. Volchenkov, L. Volchenkova and Ph. Blanchard, Phys. Rev. E \textbf{66}%
, 046137 (2002).

[9]R. Pastor-Satorras and A. Vespignani, Phys. Rev. E \textbf{65}, 036104
(2002).

[10]Lazaros K. Gallos and Panos Argyrakis, Physica A \textbf{330}, (2003)
117.

[11]Zanette, Dami\'{a}n H., Kuperman, and Marcelo, Physica A \textbf{309},
(2002) 445.

[12]F.G. Ball, Threshold behaviour in stochastic epidemics among households,
in: C.C. Heyde,Y.V. Prohorov, R. Pyke and S.T. Rachev (eds.), Athens
Conference on Applied Probability and Time Series, vol. I, Applied
Probability, Lecture Notes in Statistics \textbf{114}, 253 (1996).

[13]F.G. Ball, D. Mollison, and\ G. Scalia-Tomba, Epidemics with two levels
of mixing, Ann. Appl. Probab. \textbf{7}, 46 (1997).

[14]N.G. Becker and R. Hall, Immunization levels for preventing epidemics in
a community of households made up of individuals of di.erent types, Math.
Biosci. \textbf{132}, 205 (1996).

[15]N.G. Becker and D.N. Starczak, Optimal vaccination strategies for a
community of households, Math. Biosci. \textbf{139}, 117 (1997).

[16]N.G. Becker, A. Bahrampour, and K. Dietz, Threshold parameters for
epidemics in different community settings, Math. Biosci. \textbf{129}, 189
(1995).

[17]N.G. Becker and K. Dietz, The effect of the household distribution on
transmission and control of highly infectious diseases, Math. Biosci. 
\textbf{127}, 207 (1995).

[18]F. Ball, Math. Biosci. \textbf{156,} 41(1999).

[19]G. Ghoshal, L. M. Sander, and I. M. Sokolov, e-print
cond-mat/0304301(2003).

[20]Rinaldo B. Schinazi, Theoretical Population Biology \textbf{61}, 163
(2002).

[21]S.H. Strogatz, Nature (London) \textbf{410}, 268(2001).

[22]S.N. Dorogovtsev and J.F.F. Mendes, Adv. Phys. \textbf{51}, 1079(2002).

[23]R. Albert and A.-L. Barab\'{a}si, Rev. Mod. Phys. \textbf{74}, 47 (2002).

[24]D.J Watts and S.H. Strogatz, Nature (London) \textbf{393}, 440 (1998).

[25]A.-L. Barab\'{a}si and R. Albert, Science \textbf{286}, 509 (1999).

[26]V\'{\i}ctor M. Egu\'{\i}luz and Konstantin Klemm, Phys. Rev. Lett. 
\textbf{89}, 108701 (2002).\newpage 

\begin{center}
Captions of Figures
\end{center}

Fig\ref{fig1} The fraction of the infected individuals, $\rho ,$as a
function of the spreading rate $\lambda $ for household structure SIS model
on regular networks with $k_{c}=4$, $N=4$. The simulations have been
averaged over 200 different realizations.

Fig\ref{fig2} The fraction of the infected individuals, $\rho ,$as a
function of the spreading $\lambda $ for household structure SIS model on
scale-free networks with $\left\langle k\right\rangle =6$, $N=4$. The
simulations have been run in networks with $10^{5}$ nodes.

\newpage

\bigskip

\bigskip

\FRAME{ftbpF}{4.075in}{3.1479in}{0pt}{}{\Qlb{fig1}}{fig1.eps}{\special%
{language "Scientific Word";type "GRAPHIC";maintain-aspect-ratio
TRUE;display "USEDEF";valid_file "F";width 4.075in;height 3.1479in;depth
0pt;original-width 4.0248in;original-height 3.1021in;cropleft "0";croptop
"1";cropright "1";cropbottom "0";filename '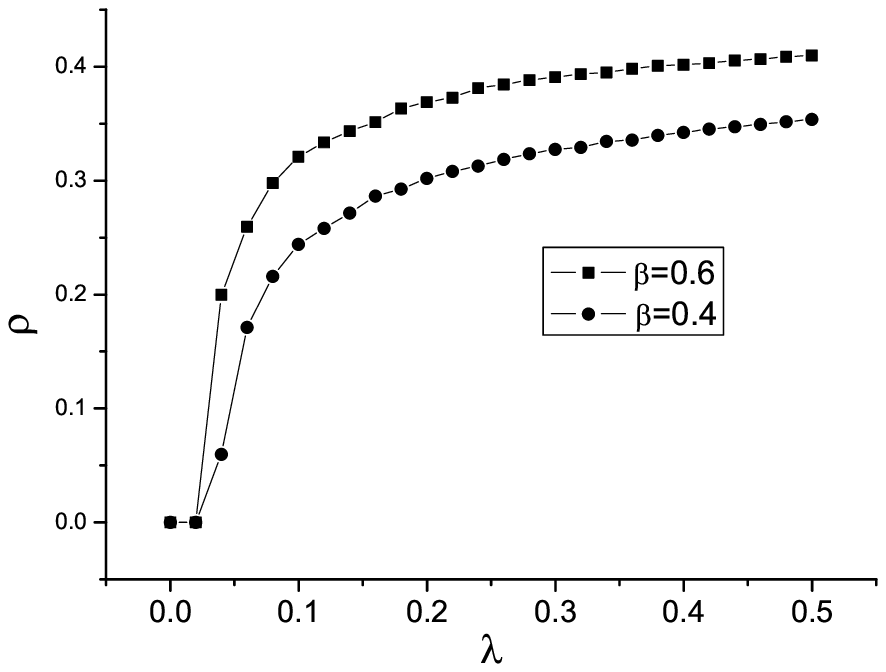';file-properties
"XNPEU";}}

\FRAME{ftbpF}{4.2557in}{3.3719in}{0pt}{}{\Qlb{fig2}}{fig2.eps}{\special%
{language "Scientific Word";type "GRAPHIC";maintain-aspect-ratio
TRUE;display "USEDEF";valid_file "F";width 4.2557in;height 3.3719in;depth
0pt;original-width 4.2056in;original-height 3.3252in;cropleft "0";croptop
"1";cropright "1";cropbottom "0";filename '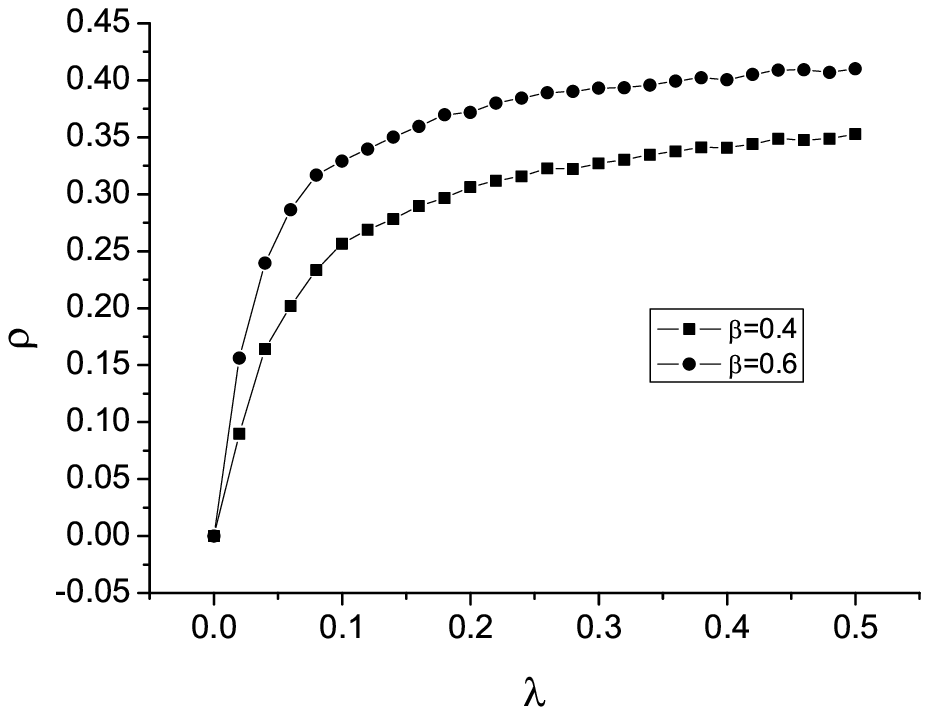';file-properties
"XNPEU";}}

\end{document}